\documentclass{emulateapj}

\usepackage{epstopdf}
\usepackage{longtable}
\usepackage{rotate}
\usepackage{amsmath}
\usepackage{lscape}

\newcommand{\goes}{{\it GOES} }
\newcommand{\goess}{{\it GOES}}
\newcommand{\xrs}{{\it GOES}/XRS }
\newcommand{\xrss}{{\it GOES}/XRS}

\newcommand{\rhessis}{{\it RHESSI}}

\newcommand{\sdos}{{\it SDO}}
\newcommand{\eve}{{\it SDO}/EVE }
\newcommand{\eves}{{\it SDO}/EVE}
\newcommand{\aia}{{\it SDO}/AIA }

\newcommand{\xrt}{{\it Hinode}/XRT }
\newcommand{\xrts}{{\it Hinode}/XRT}

\newcommand{\specialcell}[2][c]{\begin{tabular}[#1]{@{}c@{}}#2\end{tabular}}

\begin{document}

\title{DECAY PHASE COOLING AND INFERRED HEATING OF M- AND X-CLASS SOLAR FLARES}

\author{Daniel F. Ryan\altaffilmark{1,2,3}, Phillip C. Chamberlin\altaffilmark{2}, Ryan O. Milligan\altaffilmark{4,2,3}, Peter T. Gallagher\altaffilmark{1}}

\altaffiltext{1}{School of Physics, Trinity College Dublin, Dublin 2, Ireland.}
\altaffiltext{2}{Solar Physics Laboratory (Code 671), Heliophysics Science Division, NASA Goddard Space Flight Center, Greenbelt, MD 20771, U.S.A.}
\altaffiltext{3}{Catholic University of America, Washington, DC 20064, U.S.A.}
\altaffiltext{4}{Queen's University Belfast, Belfast BT7 1NN, Northern Ireland.}

\begin{abstract}
In this paper, the cooling of 72 M- and X-class flares is examined using \xrs and \eves.  The observed cooling rates are quantified and the observed total cooling times are compared to the predictions of an analytical 0-D hydrodynamic model.  It is found that the model does not fit the observations well, but does provide a well defined lower limit on a flare's total cooling time.  The discrepancy between observations and the model is then assumed to be primarily due to heating during the decay phase.  The decay phase heating necessary to account for the discrepancy is quantified and found be $\sim$50\% of the total thermally radiated energy as calculated with \goess.  This decay phase heating is found to scale with the observed peak thermal energy.  It is predicted that approximating the total thermal energy from the peak is minimally affected by the decay phase heating in small flares.  However, in the most energetic flares the decay phase heating inferred from the model can be several times greater than the peak thermal energy.
\end{abstract}

\section{Introduction}
\label{sec:intro}
Solar flares are among the most powerful events in the solar system, releasing up to $10^{32}$ ergs in a few hours or even minutes.  They are believed to be powered by magnetic reconnection, a process whereby energy stored in coronal magnetic fields is suddenly released.  This causes a rapid heating and expansion of the flare plasma which is then believed to cool by conductive, radiative and enthalpy-based processes.  However, the balance between cooling and heating in solar flares and the processes which determine this are still not fully understood.  A greater insight into this interaction would allow us to better constrain the energy release mechanisms of solar flares.

To date there have been many studies aimed at modeling the heating and cooling of solar flares \citep[e.g.][]{moor75,anti78, anti80,fish85, dosc83, carg93, klim01, reev02, brad05, klim06, warr06, warr07, sark08}.  These include full 3-D magnetohydrodynamic (MHD) models as well as 1-D MHD models.  1-D models assume that flare loop strands are magnetically isolated and therefore only solve the MHD equations along the axis of the magnetic field \citep[e.g.][]{brad05}.  This is less computationally draining than the full 3-D treatment and therefore allows a higher resolution, more useful for detailed comparison with observation.  0-D models have also been developed \citep[e.g.\ Enthalpy-Based Thermal Evolution of Loops, EBTEL;][]{klim08} which treat field aligned average properties.  Although these models sacrifice some completeness, they are much faster to run allowing an easier exploration of the dependence of results on different possible coronal property values.  Although these models are valuable in increasing our understanding of flare heating and cooling they nonetheless suffer from drawbacks. These can include arbitrary inputs of unobservable parameters such as heating function and number of loop strands.

Previous studies focused on observations of flare cooling are also numerous.  \citet{culh70} compared simple collisional, radiative, and conductive cooling models to observations of four flares made with the fourth {\it Orbiting Solar Observatory} (OSO-4).  They found that collisional cooling was unphysical while conduction and radiation were equally plausible.  Although they could not determine which was dominant, they did find that that for radiative cooling to dominate, the flare density would have to be high ($\gtrsim$10$^{11}$~cm$^{-3}$).  In contrast, conduction would require low densities ($\sim$10$^{10}$~cm$^{-3}$) to dominate.

In contrast, \citet{with78} was able to compare the relative importance of cooling mechanisms.  This study examined the differential emission measure (DEM) of a single flare using {\it Skylab} and hence determined that conductive and radiative losses were comparable.  This suggests that both processes were equally important in the cooling of that flare.  From discrepancies between observations and conductive and radiative cooling models, it was determined that $\sim$10$^{31}$~ergs of additional heating must have been deposited after the flare peak.

More recently, \citet{jian06} examined loop top sources in 6 flares using the {\it Reuven Ramaty High Energy Solar Spectroscopic Imager} \citep[\rhessis;][]{lin02}.  They found that the observed cooling rate was slightly higher than expected from radiative cooling, but significantly lower than that expected from conduction.  To account for this, they calculated that more than 10$^{30}$~ergs of additional heating during the decay phase was necessary.  This was greater than that seen during the impulsive phase.  However they concluded that much of this discrepancy was more plausibly explained by suppressed conduction.

\citet{raft09} also used RHESSI along with several other instruments to chart the thermal evolution of a single C1.0 flare.  They performed a best fit to the observations using the EBTEL model and an assumed heating function to infer radiative and conductive cooling profiles.  Conduction was found to dominate initially while radiation dominated in the latter phases.

A common aspect of many flare cooling studies such as those mentioned above is that they focus on single or small numbers of events.  This means they cannot say if their findings are anomalous or characteristic of flares.  As a result, it is still unclear just how well cooling models describe ensembles of flares.  In this paper we aim to improve upon previous studies by observing the cooling profiles of 72 M- and X-class flares.  This is done using observations from X-Ray Sensors onboard the {\it Geostationary Operational Environmental Satellites} \citep[\xrss;][]{hans96} and the Extreme ultraviolet Variability Experiment onboard {\it Solar Dynamics Observatory} \citep[\eves;][]{wood12}.  The observed cooling times are then compared to predictions made by the model of \citet{carg95}, a simple, analytical 0-D model.  Although this model is highly simplified, it was chosen as a first step because it is quick and easy to apply to many flares.  In Section~\ref{sec:data} of this paper we describe our observations.  In Section~\ref{sec:model} we discuss the assumptions, limitations and equations of the \citet{carg95} model and describe how we observationally calculated the required inputs.  In Section~\ref{sec:results} we compare the observed cooling times to those predicted by the model and quantify the discrepancy.  We then infer the decay phase heating required to account for this difference.  Finally we outline our conclusions in Section~\ref{sec:conc}.

\section{Observations \& Data Analysis}
\label{sec:data}
\subsection{Instrumentation \& Flare Sample}
\label{sec:inst}
Observations for this study were taken from three instruments: the XRS onboard the \goess-14 and 15 satellites; the Multiple Extreme ultraviolet Grating Spectrograph A channel \citep[MEGS A;][]{hock12a} onboard \eves; and the X-Ray Telescope onboard {\it Hinode} \citep[\xrts;][]{golu07,kano08}.

The \xrs measures spatially integrated solar X-ray flux in two wavelength bands (long; 1--8~\AA, and short; 0.5--4~\AA) every two seconds.  Temperature can be calculated from the ratio of these channels using the method of \citet{whit05}.  In this method, coronal abundances \citep{feld92}, the ionization equilibria of \cite{mazz98}, and a constant density of 10$^{10}$~cm$^{-3}$ were assumed.  Although this final assumption is probably not true, it was justified by \citet{whit05} who used CHIANTI to compute the spectrum of an isothermal plasma at 10~MK with densities of $10^9$, $10^{10}$, and $10^{11}$ cm$^{-3}$.  No significant differences were found.

MEGS A onboard \sdos/EVE measures spatially integrated solar irradiance from 6 to 37~nm with a spectral resolution of 0.1~nm.  MEGS A is an 80$^o$ grazing incidence off-Rowland circle spectrograph and has a time cadence of 10~s.  Within its spectral range are a number of temperature and density sensitive Fe lines useful for examining the thermodynamics of hot flares.

\xrt is a grazing incidence X-ray telescope with a spatial resolution of 1~arcsec.  It provides broadband images of the Sun in wavelengths of 0.2--20~nm.  It has a maximum field of view of 34$\times$34~arcsecs but can also focus on several smaller regions of interest simultaneously.  The time cadence depends on the observing program used but is typically on the order of seconds.  \xrt has numerous filters which can help to reduce saturation during flares.  These have quite wide temperature responses, but all peak around 8--13~MK.  The temperature sensitivity, spatial resolution and time cadence of \xrt make it the most ideal instrument available for directly measuring loop lengths of hot ($>$5~MK) X-ray- and EUV-emitting flare plasma.  It is better suited than the Atmospheric Imaging Assembly \citep[AIA;][]{leme12} onboard \sdos, which is often saturated by M- and X-class flares and which has greater sensitivity to cooler coronal plasma ($\sim$1~MK). However, of the 72 flares included in this study, only 22 were well observed by \xrts.  Therefore, loop lengths were determined using the RTV-scaling law \citep{rtv78} while \xrt was used to quantify the uncertainty of this law.  See Section~\ref{sec:carginp} for further details.

The 72 M- and X-class flares examined in this study were chosen via two criteria.  Firstly, their decay phases had to be temporally isolated from other flares.  This was determined from visual inspection of the \goes lightcurves.  Secondly, the flares had to be observed to cool to at least 8~MK with either the \xrs or \eve MEGS-A.  A complete list of the flares and their properties are listed in Table~2 in the Appendix.

\subsection{Observing Flare Cooling}
\label{sec:cool}

\begin{table}
\begin{center}
\caption[Wavelengths and temperatures of bandpasses and emission lines used in measuring cooling rates]{Wavelengths and temperatures of bandpasses and emission lines used in measuring cooling rates}
\label{tab:lines}
\begin{tabular}{cccc}
\hline
\hline
Instrument	&Wavelength [nm]	&Temperature [MK]		\\
\hline
\xrs			&\specialcell{0.05--0.4 -- Short\\0.1--0.8 -- Long}	&$>$4	\\
\hline
Ion			&Wavelength [nm]	&Temperature [MK]	\\
\hline
Fe XXIV		&19.20			&15.8			\\
Fe XXII		&11.71			&12.6			\\
Fe XIX		&10.83			&10.0			\\
Fe XVIII		&9.39			&7.9				\\
Fe XVI		&33.54			&6.3				\\
Fe XV		&28.41			&2.5				\\
Fe XIV		&26.47			&2.0				\\
\hline
\end{tabular}
\end{center}
\end{table}

\begin{figure}
\begin{center}
\includegraphics[height=16cm]{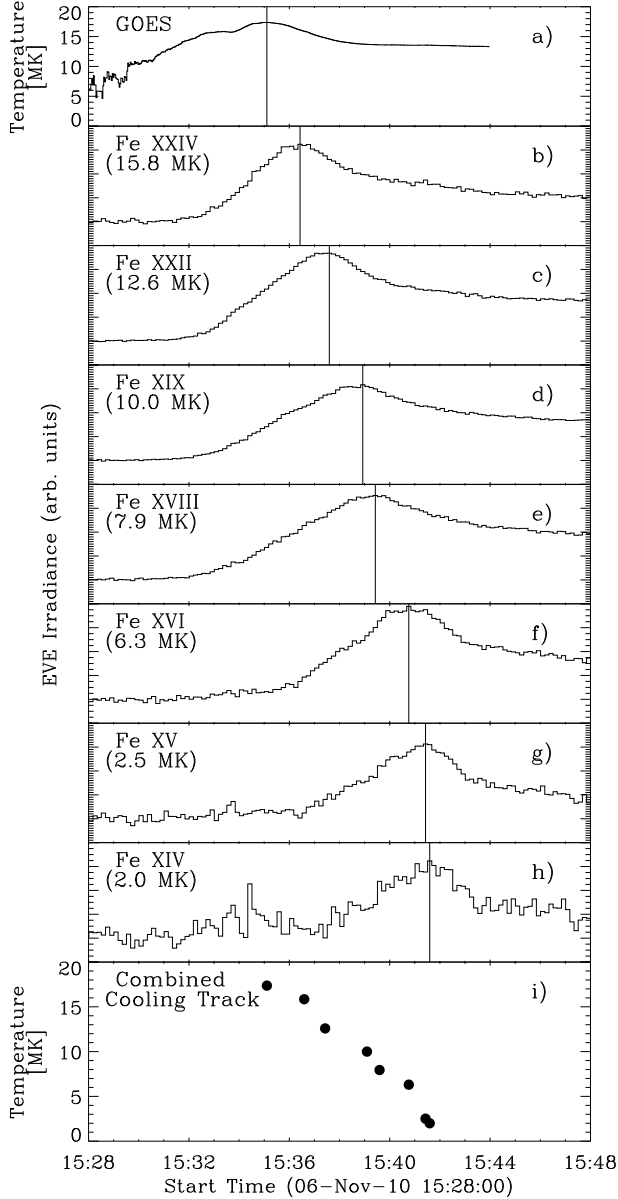}
\caption{Cooling track for the 2010-Nov-06 M5.5 flare which began at 15:28~UT. {\it a)} Background-subtracted \goes temperature profile. Peak is marked by the vertical line.  {\it b) -- h)} Lightcurves of sequentially cooler Fe lines ranging from 15.8~MK to 2~MK observed by \eve MEGS-A. The peak of each lightcurve is also marked by a vertical line.  {\it i)} Combined cooling track obtained by plotting the time of the peak of each profile (including \goes temperature profile) with its associated peak temperature.  The resultant cooling time is the duration of this cooling track.}
\label{fig:coolprofs}
\end{center}
\end{figure}

The cooling of the flares in this study was charted by combining the peak of the \goes temperature profile with the peaks of lightcurves of various temperature sensitive Fe lines observed by \eve MEGS-A.  The \goes temperature was calculated using the TEBBS method \citep[Temperature and Emission measure-Based Background Subtraction;][]{ryan12} which performs an automatic background subtraction and finds the temperature and emission measure using the method of \citet{whit05}.  The Fe lines used in this study along with their formation temperatures are listed in Table~\ref{tab:lines}.  These lines were chosen because in the conditions of a solar flare, they are dominant over neighbouring lines within the MEGS-A resolution and therefore minimally blended.  Before extracting these lightcurves, a background subtraction was made to each observed flare spectrum.  The background spectrum was found by averaging the spectra within a quiet period before the flare start time.  This period was determined for each flare by visual inspection of the \goes lightcurves.  This helped ensure that the behaviour of the lightcurves was minimally contaminated by emission from non-flaring plasma.  The irradiance observed at the wavelength of each line in Table~\ref{tab:lines} was then summed with that within $\pm$0.05~nm, i.e.\ the spectral resolution of MEGS-A.  This was done for each spectrum taken during the flare and hence flare lightcurves were formed.  A cooling track was then generated by plotting the peak time of each lightcurve against its associated formation temperature.  The cooling time was then given by the duration of this track.

\begin{figure}
\begin{center}
\includegraphics[height=12cm]{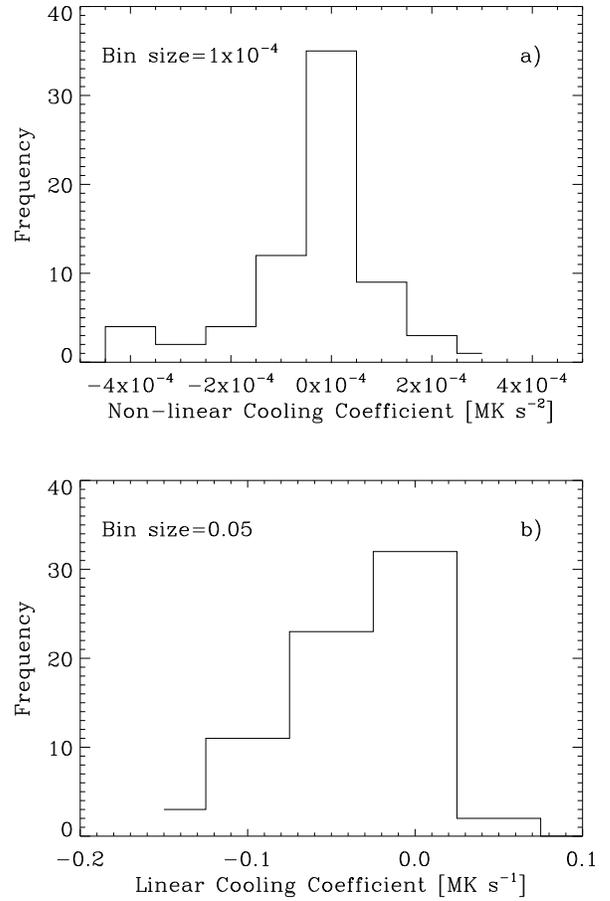}
\caption{Histograms showing the non-linear (panel a) and linear (panel b) coefficients of the second order polynomial fits to the observed cooling profiles of the 72 M- and X-class flares in this study.}
\label{fig:coolfithist}
\end{center}
\end{figure}

Figure~\ref{fig:coolprofs} shows an example for an M5.5 flare which occurred on 2010 November 06 at 15:27~UT.  Figure~\ref{fig:coolprofs}{\it a} shows the \goes temperature curve while Figures~\ref{fig:coolprofs}{\it b}--\ref{fig:coolprofs}{\it h} show the lightcurves of the Fe lines measured by \eves.  The vertical lines in each panel mark the peak time of that lightcurve.  The lightcurves peak in order of descending temperature.  This is interpreted as being due to the plasma cooling.  Figure~\ref{fig:coolprofs}{\it i} shows the resulting cooling track, with each datum point representing the peak time and temperature associated with the lightcurves above.  From this it can be seen that this flare cooled from 17~MK to 2~MK over the course of 389~$\pm$~10~seconds.  The uncertainty comes from combining the time resolutions of the \xrs and \eve in quadrature.

In order to parameterize the flare cooling, each flare's cooling profile was fit with a second order polynomial of the form
\begin{equation}\label{eqn:tobsfit}
T(t) = T_0 + \theta t + \mu t^2 \mbox{       MK}
\end{equation}
where $t$ is time since the start of the cooling phase in seconds, $T_0$ is the temperature at the start of the cooling phase in megakelvin (i.e.\ \goes peak temperature), $\theta$ is the linear cooling coefficient [MK s$^{-1}$], and $\mu$ is the non-linear cooling coefficient [MK s$^{-2}$].  Figure~\ref{fig:coolfithist}{\it a} shows a histogram of the non-linear cooling coefficients, $\mu$.  The distribution is very narrowly peaked around zero with a full width half max of $\lesssim$10$^{-4}$~MK s$^{-2}$ and a mean of -6$\times$10$^{-5}$~MK s$^{-2}$.  This implies that the majority of the flare cooling profiles are very linear.   This agrees qualitatively \citet{raft09}, whose observed cooling profile of a C1.0 flare was also quite linear.  Figure~\ref{fig:coolfithist}{\it b} shows a histogram of the linear cooling coefficients, $\theta$.  Since the non-linear cooling coefficients are so small, the linear cooling coefficients approximate the cooling rates.  The histogram ranges from -1.5--0~MK~s$^{-1}$ and has a mean of -0.035~MK~s$^{-1}$.  This implies that the SXR-emitting plasma of an average M- or X-class flare cools at a rate of $\sim$3.5$\times$10$^{4}$~K~s$^{-1}$.  It should be noted that although the histogram in Figure~\ref{fig:coolfithist}{\it b} peaks at the bin centered on zero, all flares have non-zero linear cooling coefficients.  These parameterizations are used again in Section~\ref{sec:decayenergy}.

\section{Modelling}
\label{sec:model}
\subsection{The Cargill Model}
\label{sec:theory}
To model the cooling observations discussed in the previous section, we used the model of \citet{carg95}.  This model is based on conductive and radiative cooling timescales derived from the energy transport equation which describes how a plasma's thermal energy density changes with time.  To derive these timescales, a number of assumptions are made.  First, the plasma is isotropic, i.e.\ contains no shearing motions.  Second, it is isothermal, i.e.\ obeys the ideal gas law, $p = N k_B T$.  Third, the plasma $\beta$ is low, implying all particle motions are along the axis of the magnetic field.  Fourth, the plasma is `collisionless'.  Fifth, the conductive heat flux obeys Spitzer conductivity, $\kappa_0 T^{5/2} \nabla T$, where $\kappa_0 = 10^{-6}$.  And sixth, the radiative loss function, $P_{rad}$, between 10$^6$ and 10$^7$~K is adequately modeled by $P_{rad} = \chi N^{\zeta} T^{\alpha}$~erg~cm$^3$~s$^{-1}$, where $\chi = 1.2 \times 10^{-19}$, $\zeta = 2$ and $\alpha = -0.5$ \citep{rtv78}.  By using these assumptions, the energy transport equation can be written as
\begin{multline}\label{eqn:et_q}
\frac{1}{\gamma - 1} \frac{\partial p}{\partial t} = - \frac{1}{\gamma - 1}  \frac{\partial}{\partial s}( p u_s ) - p \frac{\partial u_s}{\partial s} - \\ \left( \frac{\partial}{\partial s} \left[ \kappa_0 T^{5/2} \frac{\partial T}{\partial s} \right] - \chi N^\zeta T^{\alpha} \right) + h
\end{multline}
where $p$ is pressure, $s$ is the spatial coordinate along the axis of the magnetic field, $u_s$ is the plasma velocity along the axis of the magnetic field, $T$ is temperature, $N$ is density, and $h$ is the heat energy added per unit volume, per unit  time.

In order to derive the characteristic timescales of the cooling mechanisms, \citet{carg95} assumed that there were no flows within the plasma ($u_s = 0$) and that there was no heating ($h = 0$).  This implies that the only way the thermal energy density is altered is via conductive and radiative heat flux.  Then the characteristic conductive cooling timescale can be derived by neglecting radiative processes in Equation~\ref{eqn:et_q}, and vice versa for the characteristic radiative timescale.  By further assuming that the plasma is monatomic (i.e.\ the adiabatic constant, $\gamma = 5/3$) the conductive and radiative timescales are given by
\begin{equation}\label{eqn:tauc}
\tau_c = 4 \times 10^{-10} \frac{N L^2}{T^{5/2}}
\end{equation}
and
\begin{equation}\label{eqn:taur}
\tau_r = 3.45 \times 10^3 \frac{ T^{3/2}}{N}
\end{equation}
respectively, where $L$ is loop half length.

From these timescales, a total cooling time can be calculated if it is assumed that the flare only cools by either conduction or radiation at any one time.  If the conductive timescale is initially shorter than the radiative timescale, the flare is assumed to cool purely by conduction from its initial temperature, $T_0$, until time, $t_*$ and temperature, $T_*$, when the two timescales become equal.  From then the flare is assumed to cool purely radiatively to the final temperature, $T_L$, which takes additional time, $t_{**}$.  It should be noted here that $t_*$ and $t_{**}$ and different from $\tau_c$ and $\tau_r$.  The former are the periods when the flare cools by conduction and radiation respectively.  The latter are characteristic timescales of the cooling processes.  With this in mind, the total cooling time of the flare is given by 
\begin{multline}\label{eqn:tcandr2}
t_{tot} = \tau_{c0} \left[ \left( \frac{\tau_{r0}}{\tau_{c0}} \right)^{7/12} - 1 \right] + \\ \frac{2 \tau_{r0}}{3} \left( \frac{\tau_{c0}}{\tau_{r0}} \right)^{5/12} \left[ 1 - \left( \frac{\tau_{c0}}{\tau_{r0}} \right)^{1/6}  \left( \frac{T_L}{T_0} \right) \right]
\end{multline}
where subscript `0' implies the value of that property at the start of the cooling phase.  If conduction never dominates radiation, the flare is assumed to cool purely radiatively.  This gives a total flare cooling time of
\begin{equation}\label{eqn:tronly2}
t_{tot} = \frac{2 \tau_{r0}}{3} \left(1 - \frac{T_L}{T_0} \right) 
\end{equation}
Equations~\ref{eqn:tcandr2} and \ref{eqn:tronly2} use the relations describing the temporal evolution of temperature due to conduction derived by \citet{anti78} and due to radiation derived by \citet{anti80}.

The Cargill model is a very simple, easy-to-use analytical model.  However, its simplicity gives rise to a number of limitations.  It assumes that at any one time cooling occurs via either conduction or radiation, with an instant switch between the two when their cooling timescales are equal.  This assumption may be an acceptable approximation when either the conductive or radiative timescales are much longer than the other.  However, it is certainly not valid when the two timescales are similar.  In addition, this model does not account for enthalpy-based cooling.  This type of cooling is most significant towards the end of a flare when the temperature is low and no longer supports the plasma against gravity.  Therefore these equations are not suitable for modeling plasma cooling below $\sim$1--2~MK.

Furthermore, the Cargill model treats the flare as a monolithic loop.  Other 0-D models \citep[e.g.][]{warr06,klim08} employ the idea that flaring loops are comprised of many smaller strands, each heated and cooled at different times.  However, some recent studies \citep[e.g.][]{asch11b,pete13} have examined coronal loop cross-sections at high resolution and found no discernible structure.  This implies the such strands are below a resolution of 0.2~arcsec or that a monolithic treatment may be justified.  If the flaring loops are indeed made of sub-strands they would have the same orientation.  And since the plasma is frozen onto the magnetic field lines and diffusion across them is minimal, reducing the multiple strands to one spatial dimension along the axis of the magnetic field, as in the Cargill model, is somewhat justified via symmetry.  One could argue something similar for multiple loops in a flaring arcade.  However, this treatment implicitly assumes that all the loops/strands are being heated and cooled simultaneously which results in the average behavior of all the loops/strands being modeled.  Despite this restriction, such an approach can still be useful in examining flare hydrodynamics, especially since the additional free parameters introduced by multi-strand modeling are very unwieldy when modeling a large number of flares.

Finally, the Cargill model does not account for heating which has been suggested can continue well into the decay phase \citep[e.g.][]{with78,jian06,warr06}.  Thus this assumption is not very well justified and would be expected to produce shorter predicted cooling times than those observed.  Despite the above limitations, the Cargill model contains much of the physics believed to be responsible for flare cooling and quantifying how well it simulates observations is important to better understand flare evolution.

\begin{figure*}
\begin{center}
\includegraphics[height=20cm]{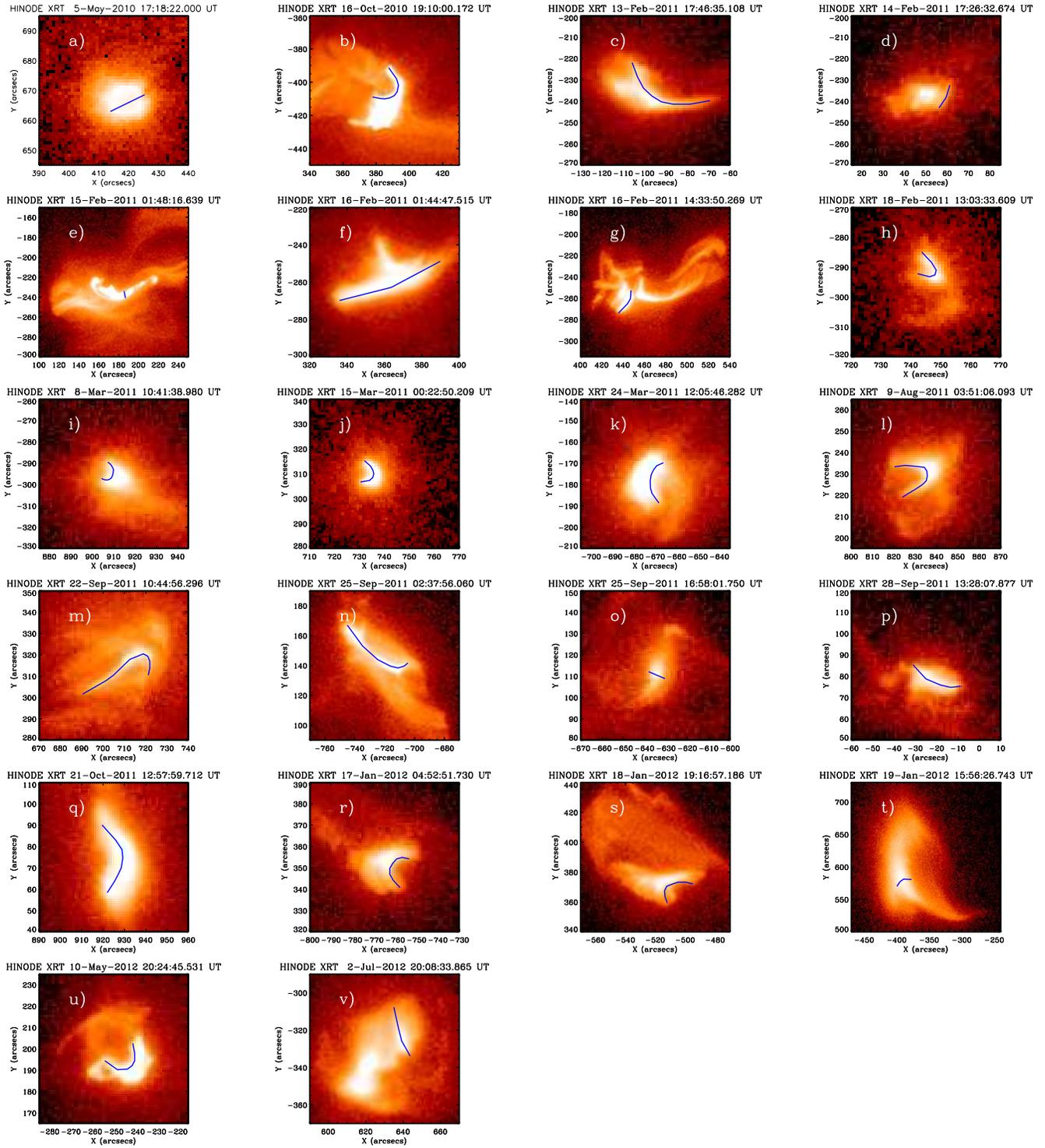}
\caption{Observed loop lengths measured with \xrt of 22 flares.}
\label{fig:xrt}
\end{center}
\end{figure*}

\subsection{Observed Inputs to Cargill Model}
\label{sec:carginp}
The observed inputs required by the Cargill model are, initial temperature, initial density, and loop half length which is assumed to be constant.

The initial temperature, $T_0$, is that at the beginning of the observed cooling track i.e.\ the \goes temperature peak.  As stated in Section~\ref{sec:cool} this was calculated using the TEBBS method \citep{ryan12}.  This measurement has two sources of uncertainty: one due to the instrument \citep[][Section 7]{garc94} and one due to the background subtraction \citep[][Section 3.2]{ryan12}.  The total uncertainty in the initial temperature was found by combining these two uncertainties in quadrature.

The initial density, $N_0$, was determined using the method of \citet{mill12}.  This method uses CHIANTI \citep[version 7;][]{land12} to determine the relationship between density and the ratios of three density dependent Fe XXI line pairs (12.121~nm/12.875~nm, (14.214~nm~+~14.228~nm)/12.875~nm, and 14.573~nm/12.875~nm).  In this study, only the first ratio was used as only these lines consistently exhibited increased emission due to the flares.  This method is only valid for temperatures above 10~MK due to the formation temperatures of these lines.  It is also not sensitive outside the range 10$^{10}$--10$^{14}$~cm$^{-3}$.  However, the Cargill model only requires the initial density, i.e.\ the density at the time of the peak \goes temperature.  Since all the flares in this study peak above 10~MK and were found to have densities within this range, the method of \citet{mill12} is suitable.

The uncertainty in the density measurement was taken from the uncertainty in the line intensity ratio as measured by EVE.  The ratio itself has two main sources of uncertainty.  First, the instrumental uncertainty of the irradiance of the two lines.  Second is the uncertainty due to the noise in the ratio time profile.  The latter was evaluated as the standard deviation during the time period when the flare was hotter than 12~MK as determined with the \xrss.  This threshold was chosen as it ensured that the flare temperature (accounting for uncertainty) was in the valid range of the \citet{mill12} method.  The total uncertainty in the ratio was determined from the standard propagation of errors of the two uncertainty sources.  This uncertainty was then transformed to density by propagating the ratio's upper and lower limits as per \citet{mill12}.  It should be noted that there are additional uncertainties associated modelled relationship between FeXXI line intensities and density.  However, these are expected to be much smaller than the uncertainty sources discussed above.  For more information on the modelling of the FeXXI lines in CHIANTI see Section~4.7.1\ of \citet{dere97}, and references therein.

\begin{figure}
\begin{center}
\includegraphics[height=8.8cm, angle=90]{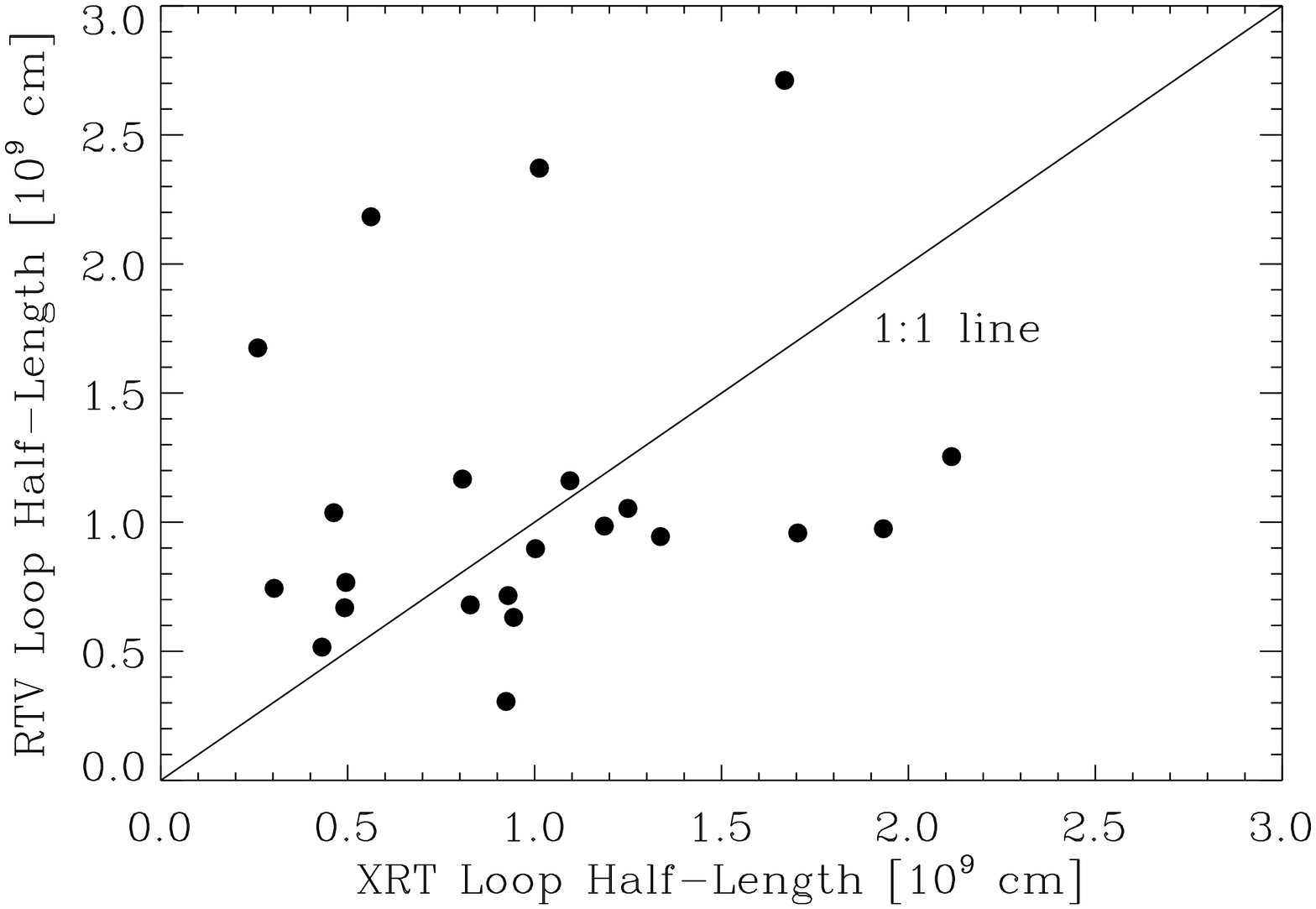}
\caption{Comparison of RTV-predicted flare loop half lengths with those measured with \xrts.  A loose correlation can be seen around the 1:1 line (over-plotted).}
\label{fig:rtvxrt}
\end{center}
\end{figure}

As stated in Section~\ref{sec:inst}, loop half length, $L$, was determined using the RTV scaling law \citep{rtv78}
\begin{equation}\label{eqn:rtv}
L = (1.4 \times 10^3)^{-3} \left( \frac{T_{max}^3}{p} \right)
\end{equation}
where $p$ is pressure and $T_{max}$ is the maximum temperature in the loop.  By assuming that the plasma is isothermal and obeys the ideal gas law, this can be rewritten in terms of temperature, $T$, and density, $N$, which can be calculated using \xrs and \eve respectively.
\begin{equation}\label{eqn:rtvtn}
L = \frac{1}{k_B (1.4 \times 10^3)^3} \frac{T^2}{N}
\end{equation}

Ideally, these loop half lengths would be directly measured using \xrts.  However, only 22 of the events in this study were well-observed by this instrument and so it was necessary to utilize the RTV-scaling.

In order to quantify the uncertainties of the RTV-predicted values, the loop lengths of these 22 events were measured with \xrt (Figure~\ref{fig:xrt}).  These are plain-of-sky measurements performed `by eye'.  A more rigorous analysis attempting to account for projection affects is expected to alter the measured loop lengths only up to a factor of $\sim$2 which is sufficient for our purposes.  The different XRT filters used for each event can be found in Table~2 in the Appendix.  These filters all peak between 8--13~MK.  Their response functions show contributions from plasma at temperatures below 1~MK of at least 2 orders of magnitude lower than the peak.  This suggests that the images are not significantly contaminated by emission from lower temperature plasma which might otherwise affect the measured loop lengths.  The one exception is the Al-mesh filter.  But this was only used for one event and was not found to be an outlier.  Where possible, \aia was used to help determine the axis of the magnetic field, along which the loop length should be measured.  For instance, a combination of AIA and XRT movies revealed that the long axis of the flares in the panels e), o), t) of Figure~\ref{fig:xrt} were flare arcades and the loop lengths themselves were actually along the shorter axis.  Despite its usefulness in these instances, AIA's sensitivity to cooler plasma and greater tendency to saturate made it less well suited to making the actual measurements than XRT.  The loop half lengths implied by the XRT measurements were compared to the RTV-predicted values (Figure~\ref{fig:rtvxrt}) and a loose correlation around the 1:1 line was found.

\begin{figure*}
\begin{center}
\includegraphics[height=13.5cm, angle=90]{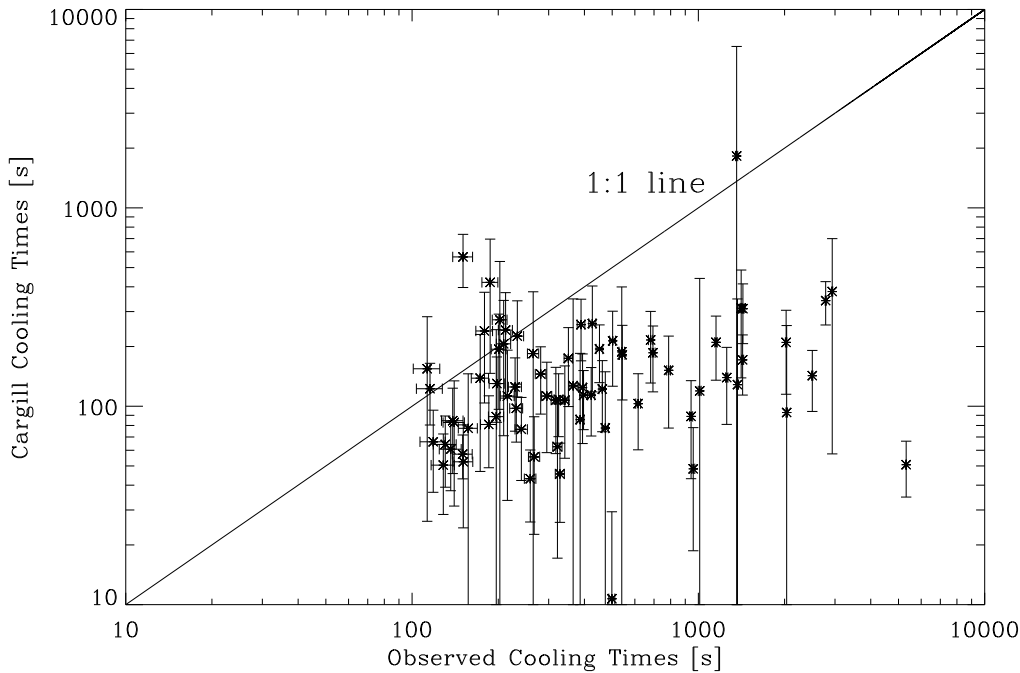}
\caption{Comparison of Cargill-predicted cooling times with observed cooling times.  The 1:1 line is overplotted for clarity.}
\label{fig:cargobs}
\end{center}
\end{figure*}

The CVRMSD (coefficient of variation of the root-mean-square deviation) of the distribution in Figure~\ref{fig:rtvxrt} was used to quantify the uncertainty of the RTV-predicted loop half lengths.  This was found to be 1.8, implying that the loop half length and uncertainty is given by $L = L_{RTV} \pm 1.8 \cdot L_{RTV}$.  This compares favourably with \citet{asch13c} who compared RTV-predicted loop lengths with the length-scales of less intense flares using \aia and found an uncertainty of $\pm1.6 \cdot L_{RTV}$.

Having measured the initial temperature, initial density and loop half length, the Cargill-predicted cooling times were found from Equation~\ref{eqn:tcandr2} if $\tau_{c0} < \tau_{r0}$ and Equation~\ref{eqn:tronly2} if $\tau_{c0} > \tau_{r0}$.  The uncertainties on these cooling times were calculated by first rewritting Equations~\ref{eqn:tcandr2} and \ref{eqn:tronly2} in terms the observed input properties (temperature, density and loop half length) and then propagating their uncertainties by the standard error propagation rules.  Having done this, we then compared the model-predicted cooling times with the observations discussed in Section~\ref{sec:cool}.

\section{Results \& Discussion}
\label{sec:results}
\subsection{Comparing Observed and Modeled Cooling Times}
\label{sec:coolcomp}
Figure~\ref{fig:cargobs} shows the comparison of Cargill-predicted and observed cooling times for 72 M- and X-class flares.  The 1:1 line is over-plotted for clarity.  It can clearly be seen that Cargill is consistent with observations at the shortest cooling times, but is not a good overall fit to the distribution.  Upon closer inspection it was found that only 14 events (20\%) had observed cooling times which agreed with Cargill within experimental error.  Meanwhile 58 (80\%) disagreed.  Of those, only 1 was over-estimated by Cargill.  The remaining 57 were underestimated.  Thus these results statistically prove that the Cargill model provides a lower limit to the time needed for a flare to cool.  In addition, it was found in 52 flares (72\%) that radiation dominated conduction for the entirety of the cooling phase.  Conduction initially dominated radiation in only 20 flares (28\%).  This suggests that flares for which radiation is the dominant cooling mechanism \citep[such as those examined by][]{mcti93,lofu07} are far more common than those in which conduction initially dominates \citep[e.g.\ ][]{moor75,jian06,raft09}.  Furthermore, \citet{culh70} concluded from examining a simple radiative cooling model that flare plasma cooling by radiation in a timescale of $\sim$500~s would exhibit high densities (10$^{11}$--10$^{12}$~cm$^{-3}$).  The average observed cooling time of the events in Figure~\ref{fig:cargobs} is 653~s and their average density is 1.4$\times$10$^{12}$~cm$^{-3}$, which is very close to the conclusions of \citet{culh70}.  This strengthens the claim that radiation is typically dominant over conduction throughout a flare's decay phase.

To further quantify the discrepancy between predicted and observed cooling times (henceforth referred to as the `excess cooling time'), the root-mean-square deviation (RMSD) of the distribution was calculated.  This was found to be 961~s.  Normalizing this to the mean of the observed cooling times (653~s) gives the coefficient of variation of the root-mean-square deviation (CVRMSD).  This quantifies the spread of the `excess cooling times' relative to the mean of the observed cooling times.  The CVRMSD was found to be 1.47 indicating a large spread as is visually suggested in Figure~\ref{fig:cargobs}.

If the Cargill model is adequately describing the cooling mechanisms of solar flares, the `excess cooling time' suggests that there is additional heating occurring throughout the decay phase.  Similar assumptions have been made in previous studies \citep[e.g.][]{with78,jian06,hock12b}.  In the following section we explore just how much additional heating energy is required to account for the `excess cooling times' and examine the distributions of these energies.

\subsection{Inferring Heating During Decay Phase}
\label{sec:decayenergy}
For radiatively dominated flares, the decay phase heating required to account for the `excess cooling time' can be determined from the following modified version of the energy transport equation
\begin{equation}\label{eqn:et_heat}
3 k_B N_0 \frac{\partial T}{\partial t} = - \chi N_0^{\zeta} T^{\alpha} + h
\end{equation}
where $k_B$ is Boltzmann's constant, $N_0$ is the density (assumed to be constant and equal to the initial density to remain consistent with the Cargill model), $T$ is temperature, $t$ is time, $\chi$, $\zeta$ and $\alpha$ have the same values as above (1.2 $\times$ 10$^{-19}$, 2, and -0.5, respectively), and $h$ is the heating rate per unit volume.  This equation is stating that the rate of change of thermal energy density (LHS), is determined by the radiative energy losses (1st term, RHS) and heating (2nd term, RHS).  The total decay phase heating energy, $H$, can then be evaluated by integrating over time and multiplying by flare volume, $V$, assumed to be constant.
\begin{multline}\label{eqn:decayenergy}
H = V \times \\ \int_0^{t_{tot}} \left[ 3 k_B N_0 \frac{\partial T(t)}{\partial t} + 1.2 \times 10^{-19} N_0^2 T(t)^{-1/2} \right] dt
\end{multline}

This analysis was performed on 38 flares within our sample (marked in Table~2 in the Appendix by their non-zero values in the `Decay Phase Heating' column).  These flares were chosen because the Cargill model implied that radiation was the dominant cooling mechanism throughout their decay phases, making Equations~\ref{eqn:et_heat} and \ref{eqn:decayenergy} valid.  These flares were also seen to cool down to at least 6~MK so the majority of their cooling could be analyzed.  The rate of change of temperature, $dT/dt$, in Equation~\ref{eqn:decayenergy} was found by differentiating the second order polynomial fits to the cooling profiles discussed in Section~\ref{sec:cool}.   The flare volume was calculated from the density and peak emission measure using the equation,
\begin{equation}\label{eqn:volume}
V = \frac{EM}{N^2}
\end{equation}
The emission measure was calculated from ratio of the \goes long channel flux and temperature using the same assumptions and methods as described in Section~\ref{sec:inst} \citep[][TEBBS]{whit05,ryan12}.  The total decay phase heating required to account for the `excess cooling time' was then calculated from Equation~\ref{eqn:decayenergy}.

\begin{figure}
\begin{center}
\includegraphics[angle=90, width=8.8cm]{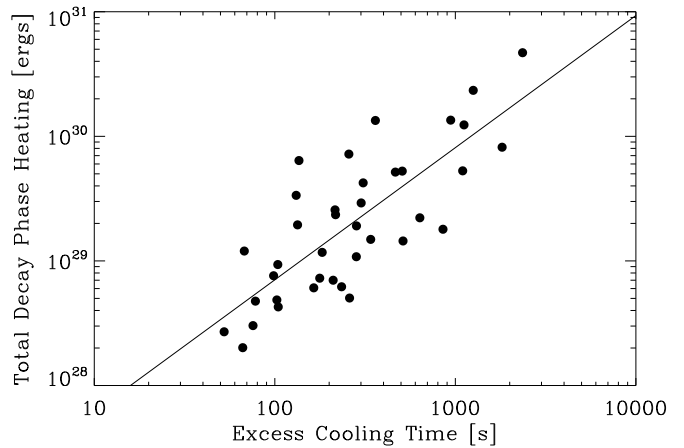}
\caption{Heating during the decay phase as a function of the difference between the observed and Cargill-predicted cooling times for  38 M- and X-class flares.  The line over-plotted is the best fit to the data (see Equation~\ref{eqn:timeenergy_fit}).}
\label{fig:timeenergy}
\end{center}
\end{figure}

\begin{figure}
\begin{center}
\includegraphics[width=8.8cm]{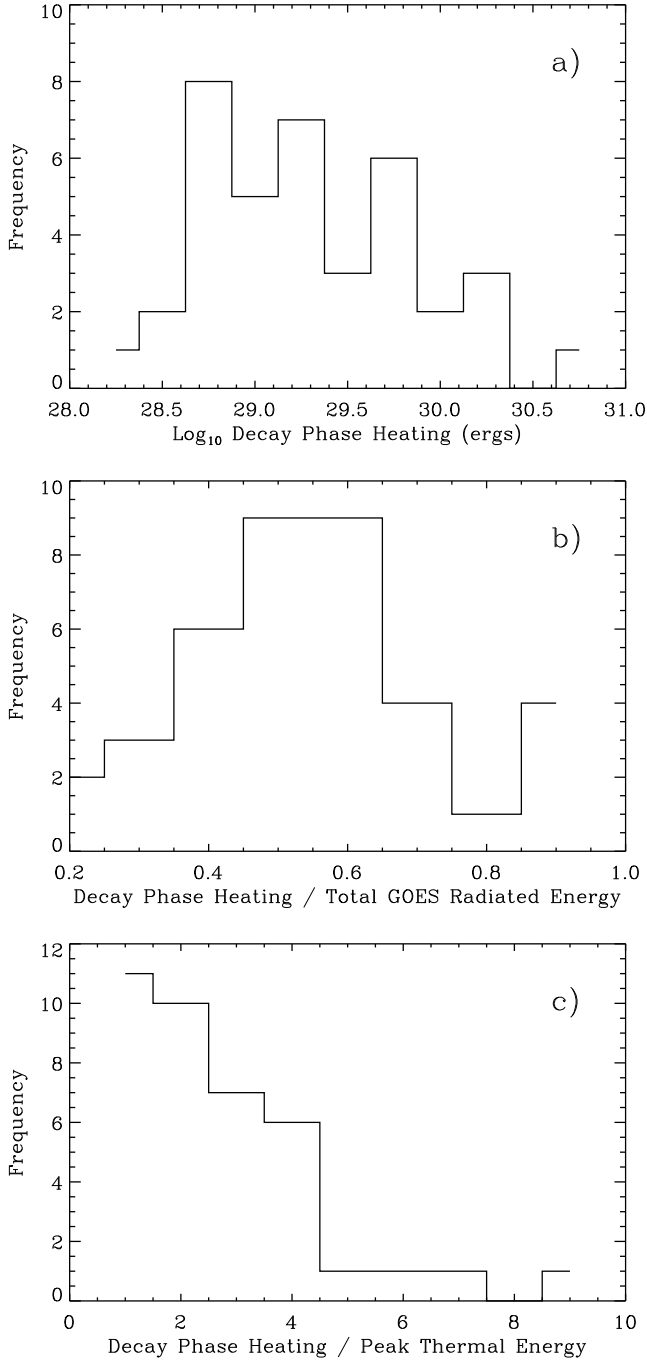}
\caption{Histograms showing the required total heating during the decay phase of 38 M- and X-class flares to account for the difference between the Cargill-predicted observed cooling times (`excess cooling time'). {\it a)} Log$_{10}$ of total decay phase heating.  {\it b)} Total decay phase heating normalized by the total energy radiated by the flare as measured by \goess.  {\it c)} Total decay phase heating divided by the thermal energy at the beginning of the cooling phase.}
\label{fig:decayenergy}
\end{center}
\end{figure}

Figure~\ref{fig:timeenergy} shows resultant energies as a function of the `excess cooling time'.  The Pearson correlation coefficient of the distribution was calculated in log-log space and found to be 0.77, implying a statistically significant correlation.  The following power-law was fit to the data,
\begin{equation}\label{eqn:timeenergy_fit}
H = 10^{26.73} (\Delta t)^{1.06 \pm 0.24}
\end{equation}
where $H$ is the total heating energy during the decay phase, and $\Delta t$ is the `excess cooling time'.  The uncertainty on the exponent represents one standard deviation.  This power-law quantifies, in very simple terms, the affect of heating during the decay phase on a flare's cooling time.

Figure~\ref{fig:decayenergy}{\it a} shows a histogram of these energies which range from 2$\times$10$^{28}$~--~5$\times$10$^{30}$~ergs.  The findings of \citet{with78} and \citet{jian06} fit into the upper limit of this range.  They inferred total decay phase heating of 10$^{31}$~ergs for the 1973 September 7 flare and $>10^{30}$~ergs in the 2002 September 20 flare, respectively.  Although the energies in Figure~\ref{fig:decayenergy}{\it a} are plausible, further testing of the Cargill model is necessary to categorically prove whether they are correct.  Nonetheless, from these heating calculations, it is possible to work out some implications of these energies in fact being correct.  This provides extra ways of testing the Cargill model's accuracy.

Firstly, the distribution in Figure~\ref{fig:decayenergy}{\it a} was fit with an an exponential using the method of maximum-likelihood, resulting in,
\begin{equation}\label{eqn:decayfit}
f(H) \propto e^{- \gamma \cdot H}
\end{equation}
where $f(H)$ is the number of events as a function of total decay phase heating energy, $H$, and $\gamma~=~1.7 ~(\pm0.3)~\times~10^{-30}$.  Again, the uncertainty represents one standard deviation.  An exponential fit was chosen because the Kolmogorov-Smirnov test \citep[][Chapter 5: Hypothesis Testing]{wall03} implied it was best suited to the data.  However energy frequency distributions of solar flares are often found to be power-laws \citep[e.g.][]{asch11} which may be due to self-organized criticality.  In this case, we cannot rule out the possibility that selection effects may have biased this distribution and including more events might reveal it to be more power-law-like.  With this in mind, a power-law was also fit to this distribution via the method of maximum-likelihood and found to be
\begin{equation}\label{eqn:decayfitpl}
f(H) \propto H^{-0.6\pm0.1}
\end{equation}

Next, the values in Figure~\ref{fig:decayenergy}{\it a} were compared to the total thermally radiated energy (Figure~\ref{fig:decayenergy}{\it b}).  CHIANTI was used to determine the spectra corresponding to the temperature and emission measure as calculated from \goess.  The total radiated energy was then found by integrating over all wavelengths and over flare duration.  The distribution in Figure~\ref{fig:decayenergy}{\it b} ranges from 0.2--0.9 and peaks at 0.5.  As previously stated, the Cargill model implies that radiation is the dominant loss mechanism for these flares.  If this is true, Figure~\ref{fig:decayenergy}{\it b} suggests that the total heating during the decay phase typically makes up half of the flare's total thermal energy budget.

The significance of the total decay phase heating is further highlighted in Figure~\ref{fig:decayenergy}{\it c} where it has been normalized by the thermal energy at the flare peak, calculated from the following equation.
\begin{equation}\label{eqn:eth}
E_{peak} = 3 N k_B T
\end{equation}
The distribution has a negative slope and ranges from $<$1 to $>$7.  This implies that the total decay phase heating energy inferred from the `excess cooling time' can be several times greater than the thermal energy at the peak.  This agrees with \citet{jian06} who found that the inferred decay phase heating in the 2002 September 20 event was greater than the energy deposited during the impulsive phase.  Such a result would be significant as previous studies \citep[e.g.\ ][]{emsl12} have used peak thermal energy as an estimate for the total thermal energy of a flare.  To quantify the relationship between decay phase heating, $H$, and peak thermal energy, $E_{peak}$, a power-law was fit to the data and found to be
\begin{equation}\label{eqn:qeth_pl}
H = 10^{-2.7\pm0.4} E_{peak}^{1.1\pm0.1}
\end{equation}
This implies that the total decay phase heating as a fraction of the peak thermal energy is greater for greater values of the peak thermal energy.  In the range explored here, the average total decay phase heating is $\sim$2.5 times the peak thermal energy.  However, this is expected to be less for less energetic flares.  Thus if the `excess cooling times' inferred from the Cargill model are to be believed, estimating a flare's total thermal energy from its peak is valid for small flares, but not for the most energetic events.

The predictions and comparisons made here all assume that the total decay phase heating inferred from the Cargill model is reasonable.  These predictions give further ways of testing the validity of the Cargill model via observations or more advanced modeling of decay phase heating.

\section{Conclusions}
\label{sec:conc}
In this paper, the cooling phases of 72 M- and X-class solar flares were examined with \xrs and \eves.  The cooling profiles as a function of time were parameterized and typically found to be very linear.  The average cooling rate was found to be $\sim$3.5$\times$10$^4$~K~s$^{-1}$.  These observations were compared to the predictions of the \citet{carg95} model.  Loop half lengths needed by this model were calculated via the RTV scaling law \citep{rtv78}.  The uncertainty on this law was quantified by comparing the predicted lengths of 22 flares within the sample with observations made by \xrts.  The loop half lengths predicted by RTV scaling law were typically within a factor of 3 of those seen in \xrts.

It was found that the Cargill model provides a well defined lower limit on flare cooling times, and the deviation from the model was quantified.  The root-mean-square deviation between the observations and the model was found to be 961~s and which was 1.47 of the mean observed cooling time.  Furthermore, the Cargill model finds that radiation is the dominant loss mechanism throughout the cooling phase for 80\% of flares.  For the remaining 20\%, Cargill finds that conduction dominates initially, before being superseded by radiation.

Next, the `excess cooling time' was assumed to be due to additional heating.  The total decay phase heating required to account for the `excess cooling time' was inferred for 38 flares within the sample.  The energies were found to be physically plausible, ranging from 2$\times$10$^{28}$ -- 5$\times$10$^{30}$ ergs.  The frequency distribution  could be described by either an exponential with an exponent of $-1.7 (\pm 0.3) \times 10^{-30}$ or a power-law with an exponent of $-0.6 \pm 0.1$.  These total decay phase heating energies were found to be highly correlated with the `excess cooling time' and was fit with a power-law with an exponent of 1.06$\pm$0.24 and a scaling factor of $10^{26.73}$.  It was also found that the total decay phase heating predicted from the Cargill model typically makes up about half of the thermally radiated energy budget of the hot flare plasma. Finally, it was determined that if the decay phase heating inferred from the Cargill model is to be believed, then peak thermal energy is an acceptable estimate for the total thermal energy of small flares.  However, this method would underestimate the thermal energy budget for the most energetic events.

In order to confirm of refute the findings inferred using the Cargill model, comparisons with direct observations of the decay phase heating must be made for an ensemble of flares.  This would further highlight the strengths and weaknesses of the Cargill model.  In addition, including more temperature sensitive lines in a similar analysis to this one or performing fits of the full EVE observed spectrum would give more comprehensive observations of the temperature and density evolution of the flare plasma.  Studies comparing similar observations with results of more advanced hydrodynamic simulations would also help us better understand the thermodynamic evolution and energetics of flare decay phases.

\vspace{5mm}

We would like to thank the anonymous referee for providing constructive feedback on this manuscript.  DFR would like to thank Arthur J.\ White, Trevor A.\ Bowen, Dr. Jim Klimchuk, Dr.\ Joel C.\ Allred and Dr. C.\ Alex Young for their helpful discussions.  He would also like to thank the Fulbright Association for funding the research.  PCC and DFR would like to acknowledge funding from the Living With a Star Targeted Research and Technology Program.  ROM is grateful to the Leverhulme Trust for financial support from grant F/00203/X, and to NASA for LWS/TR\&T grant NNX11AQ53G.

\begin{appendix}
\label{sec:app}
\end{appendix}

\bibliographystyle{apj}
\bibliography{msacc1}

\clearpage
\LongTables
\begin{landscape}

\begin{table}
\footnotesize
\caption[Events used in this study with observed and model-predicted cooling times and other thermodynamic properties]{Events used in this study with observed and model-predicted cooling times and other thermodynamic properties}
\label{tab:coolflares}
\tabcolsep=0.1cm
\begin{tabular}{cccccccccccccc}
\hline
\hline
Date &\goes &\goes &Observed &Cargill &T Peak &T Peak & T Min & T Min & Density & Loop Half & XRT Half & XRT Filter & Decay Ph. \\ 
 &Start Time &Class &Cooling [s] &Cooling [s] &[MK] &Time [s] & [MK] & Time [s] & [10$^{12}$~cm$^{-3}$] &Length [cm] &Length [cm] &  & Energy [ergs] \\
\hline
2010 May 05 & 17:13:00 & M1.3 & 212 & 241 & 21.0 & 17:17:59 & 2.5 & 17:21:31 & 0.9 & 1.0 & 0.5 & Al\_thick & (...) \\
2010 Jun 12 & 00:30:00 & M2.0 & 295 & 112 & 20.6 & 00:56:13 & 1.6 & 01:01:09 & 1.8 & 0.7 & (...) & (...) & 1.2 \\
2010 Jun 13 & 05:30:00 & M1.0 & 267 & 55 & 15.4 & 05:37:12 & 7.9 & 05:41:39 & 1.2 & 0.7 & (...) & (...) & (...) \\
2010 Oct 16 & 19:07:00 & M3.2 & 115 & 122 & 17.3 & 19:11:44 & 2.5 & 19:13:40 & 1.2 & 1.0 & 1.2 & Al\_mesh & (...) \\
2010 Nov 06 & 15:27:00 & M5.5 & 389 & 257 & 17.4 & 15:35:06 & 2.0 & 15:41:35 & 0.6 & 1.8 & (...) & (...) & 3.4 \\
2011 Feb 09 & 01:23:00 & M1.9 & 112 & 154 & 16.9 & 01:30:18 & 2.5 & 01:32:11 & 0.9 & 0.8 & (...) & (...) & (...) \\
2011 Feb 13 & 17:28:00 & M6.6 & 422 & 113 & 20.4 & 17:34:30 & 2.5 & 17:41:32 & 1.6 & 1.0 & 1.7 & Ti\_poly & 4.2 \\
2011 Feb 14 & 17:20:00 & M2.3 & 128 & 50 & 17.9 & 17:24:44  & 2.5 & 17:26:52 & 3.0 & 0.5 & 0.4 & Be\_thick & 0.5 \\
2011 Feb 15 & 01:44:00 & X2.3 & 501 & 213 & 24.5 & 01:53:20 & 2.5 & 02:01:42 & 1.2 & 1.7 & 0.3 & Be\_thin & (...) \\
2011 Feb 16 & 01:32:00 & M1.0 & 540 & 187 & 17.5 & 01:37:52 & 7.9 & 01:46:53 & 0.6 & 1.3 & 2.1 & Ti\_poly & (...) \\
2011 Feb 16 & 14:19:00 & M1.7 & 118 & 66 & 15.9 & 14:24:44 & 6.3 & 14:26:43 & 1.3 & 0.7 & 0.9 & Be\_thin & 0.3 \\
2011 Feb 18 & 12:59:00 & M1.5 & 215 & 112 & 20.1 & 13:02:28 & 2.5 & 13:06:03 & 1.6 & 2.2 & 0.6 & Be\_thick & 0.5 \\
2011 Mar 07 & 07:49:00 & M1.6 & 196 & 88 & 21.4 & 07:52:21 & 7.9 & 07:55:38 & 1.5 & 0.3 & (...) & (...) & (...) \\
2011 Mar 07 & 09:14:00 & M1.8 & 157 & 77 & 17.6 & 09:19:01 & 7.9 & 09:21:38 & 1.6 & 0.4 & (...) & (...) & (...) \\
2011 Mar 08 & 03:37:00 & M1.5 & 2025 & 209 & 13.0 & 03:46:13 & 2.5 & 04:19:58 & 0.4 & 1.7 & (...) & (...) & 8.2 \\
2011 Mar 08 & 10:35:00 & M5.4 & 386 & 85 & 20.2 & 10:41:02 & 2.5 & 10:47:28 & 2.1 & 0.7 & 0.5 & Be\_thick & 2.9 \\
2011 Mar 09 & 23:13:00 & X1.5 & 542 & 181 & 23.8 & 23:21:36 & 2.5 & 23:30:38 & 1.3 & 1.6 & (...) & (...) & 13.4 \\
2011 Mar 15 & 00:19:00 & M1.1 & 140 & 82 & 19.6 & 00:22:20 & 2.5 & 00:24:40 & 2.1 & 0.8 & 0.5 & Be\_thick & (...) \\
2011 Mar 24 & 12:01:00 & M1.0 & 341 & 107 & 16.7 & 12:06:01 & 2.5 & 12:11:43 & 1.2 & 0.7 & 0.8 & Al\_med & 0.6 \\
2011 Jul 27 & 15:48:00 & M1.1 & 959 & 48 & 14.0 & 15:59:24 & 7.9 & 16:15:23 & 1.1 & 1.0 & (...) & (...) & (...) \\
2011 Aug 03 & 04:29:00 & M1.7 & 136 & 60 & 19.7 & 04:31:39 & 6.3 & 04:33:55 & 2.3 & 0.9 & (...) & (...) & 0.3 \\
2011 Aug 04 & 03:41:00 & M9.3 & 451 & 194 & 18.0 & 03:54:14 & 2.0 & 04:01:45 & 0.8 & 1.3 & (...) & (...) & 7.2 \\
2011 Aug 09 & 03:19:00 & M2.5 & 1430 & 310 & 17.7 & 03:49:06 & 2.5 & 04:12:56 & 0.6 & 1.1 & 1.2 & Al\_thick & (...) \\
2011 Aug 09 & 07:48:00 & X7.3 & 233 & 225 & 32.5 & 08:03:23 & 1.6 & 08:07:16 & 1.8 & 1.9 & (...) & (...) & (...) \\
2011 Sep 07 & 22:32:00 & X1.8 & 281 & 145 & 21.8 & 22:37:42 & 2.5 & 22:42:24 & 1.4 & 1.0 & (...) & (...) & 6.4 \\
2011 Sep 22 & 10:29:00 & X1.4 & 2934 & 378 & 20.2 & 10:44:13 & 7.9 & 11:33:08 & 0.4 & 2.7 & 1.7 & Al\_thick & (...) \\
2011 Sep 24 & 17:19:00 & M3.1 & 498 & 10 & 19.5 & 17:22:10 & 7.9 & 17:30:28 & 10.9 & 0.3 & (...) & (...) & (...) \\
2011 Sep 25 & 02:27:00 & M4.6 & 228 & 125 & 20.1 & 02:31:54 & 2.5 & 02:35:43 & 1.4 & 1.0 & 1.9 & Al\_med + Al & 0.9 \\
2011 Sep 25 & 04:31:00 & M7.4 & 1152 & 209 & 18.2 & 04:39:56 & 2.5 & 04:59:09 & 0.7 & 2.0 & (...) & (...) & 13.5 \\
2011 Sep 25 & 15:26:00 & M3.8 & 325 & 108 & 15.7 & 15:32:44 & 2.5 & 15:38:09 & 1.1 & 0.7 & 0.3 & Be\_thick & 2.3 \\
2011 Sep 25 & 16:51:00 & M2.2 & 202 & 272 & 18.5 & 16:55:36 & 2.5 & 16:58:59 & 0.6 & 1.5 & (...) & (...) & (...) \\
2011 Sep 28 & 13:24:00 & M1.3 & 240 & 76 & 17.9 & 13:27:29 & 2.5 & 13:31:30 & 2.0 & 0.6 & 0.9 & Al\_med & 0.6 \\
2011 Oct 02 & 00:37:00 & M3.9 & 681 & 215 & 21.0 & 00:45:59 & 2.0 & 00:57:20 & 0.9 & 2.0 & (...) & (...) & 5.2 \\
2011 Oct 20 & 03:10:00 & M1.6 & 1360 & 1825 & 32.4 & 03:15:25 & 7.9 & 03:38:06 & 0.2 & 10.9 & (...) & (...) & (...) \\
\end{tabular}
\end{table}

\clearpage

\begin{table}
\footnotesize
\tabcolsep=0.1cm
\begin{tabular}{cccccccccccccc}
\hline
\hline
Date &\goes &\goes &Observed &Cargill &T Peak &T Peak & T Min & T Min & Density & Loop Half & XRT Half & XRT Filter & Decay Ph. \\ 
 &Start Time &Class &Cooling [s] &Cooling [s] &[MK] &Time [s] & [MK] & Time [s] & [10$^{12}$~cm$^{-3}$] &Length [cm] &Length [cm] &  & Energy [ergs] \\
\hline
2011 Oct 21 & 12:53:00 & M1.3 & 426 & 260 & 16.3 & 12:57:27 & 2.5 & 13:04:33 & 0.6 & 0.9 & 1.3 & Al\_med & (...) \\
2011 Oct 31 & 14:55:00 & M1.1 & 2033 & 92 & 29.0 & 15:00:45 & 7.9 & 15:34:39 & 2.8 & 1.2 & (...) & (...) & (...) \\
2011 Dec 25 & 18:11:00 & M4.1 & 197 & 130 & 18.9 & 18:15:15 & 2.5 & 18:18:33 & 1.3 & 1.2 & (...) & (...) & 1.2 \\
2011 Dec 26 & 02:13:00 & M1.5 & 787 & 151 & 17.2 & 02:22:06 & 2.5 & 02:35:13 & 0.9 & 1.3 & (...) & (...) & 2.2 \\
2011 Dec 29 & 21:43:00 & M2.0 & 393 & 124 & 18.2 & 21:48:00 & 7.9 & 21:54:34 & 0.8 & 2.1 & (...) & (...) & (...) \\
2012 Jan 17 & 04:41:00 & M1.0 & 942 & 88 & 15.1 & 04:46:06 & 6.3 & 05:01:49 & 0.9 & 1.2 & 0.8 & Al\_med & 1.8 \\
2012 Jan 18 & 19:04:00 & M1.7 & 461 & 122 & 15.9 & 19:09:08 & 2.5 & 19:16:50 & 1.0 & 1.2 & 1.1 & Al\_med & 1.5 \\
2012 Jan 19 & 13:44:00 & M3.2 & 5318 & 50 & 14.3 & 15:14:51 & 7.9 & 16:43:30 & 1.1 & 0.9 & 1.0 & Be\_thick & (...) \\
2012 Jan 23 & 03:38:00 & M8.7 & 1427 & 171 & 18.6 & 03:49:43 & 1.6 & 04:13:31 & 1.0 & 1.4 & (...) & (...) & 23.4 \\
2012 Feb 06 & 19:31:00 & M1.0 & 2783 & 340 & 12.9 & 19:42:31 & 2.0 & 20:28:55 & 0.2 & 0.7 & (...) & (...) & (...) \\
2012 Mar 05 & 02:30:00 & X1.1 & 2499 & 142 & 19.1 & 03:52:33 & 6.3 & 04:34:12 & 0.9 & 1.4 & (...) & (...) & 46.9 \\
2012 Mar 06 & 12:23:00 & M2.2 & 1369 & 128 & 18.3 & 12:36:54 & 7.9 & 12:59:43 & 0.8 & 1.4 & (...) & (...) & (...) \\
2012 Mar 06 & 21:04:00 & M1.4 & 365 & 126 & 18.0 & 21:06:18 & 7.9 & 21:12:23 & 1.0 & 0.7 & (...) & (...) & (...) \\
2012 Mar 06 & 22:49:00 & M1.0 & 321 & 62 & 19.0 & 22:52:31 & 2.5 & 22:57:53 & 2.6 & 0.5 & (...) & (...) & 0.5 \\
2012 Mar 14 & 15:08:00 & M2.8 & 397 & 113 & 16.0 & 15:17:38 & 2.0 & 15:24:15 & 1.1 & 0.9 & (...) & (...) & 1.9 \\
2012 Mar 17 & 20:32:00 & M1.4 & 264 & 184 & 19.0 & 20:37:41 & 2.0 & 20:42:06 & 1.0 & 1.1 & (...) & (...) & (...) \\
2012 Mar 23 & 19:34:00 & M1.0 & 172 & 138 & 18.5 & 19:39:35 & 7.9 & 19:42:28 & 1.0 & 0.7 & (...) & (...) & (...) \\
2012 Apr 27 & 08:15:00 & M1.1 & 616 & 103 & 14.8 & 08:22:32 & 2.5 & 08:32:48 & 1.1 & 1.3 & (...) & (...) & 1.4 \\
2012 May 07 & 14:03:00 & M1.9 & 1410 & 312 & 14.6 & 14:18:50 & 2.5 & 14:42:20 & 0.3 & 2.0 & (...) & (...) & 5.3 \\
2012 May 08 & 13:02:00 & M1.4 & 150 & 57 & 18.4 & 13:06:30 & 6.3 & 13:09:00 & 2.4 & 0.2 & (...) & (...) & (...) \\
2012 May 10 & 04:11:00 & M5.9 & 258 & 43 & 18.5 & 04:16:52 & 6.3 & 04:21:11 & 2.8 & 0.4 & (...) & (...) & 2.6 \\
2012 May 10 & 20:20:00 & M1.8 & 328 & 45 & 16.3 & 20:23:23 & 2.5 & 20:28:51 & 2.8 & 0.3 & 0.9 & Ti\_poly & 1.1 \\
2012 May 17 & 01:25:00 & M5.1 & 1256 & 139 & 15.8 & 01:38:06 & 6.3 & 01:59:03 & 0.6 & 1.7 & (...) & (...) & 12.4 \\
2012 Jun 03 & 17:48:00 & M3.4 & 231 & 97 & 14.9 & 17:54:26 & 2.0 & 17:58:17 & 1.2 & 0.7 & (...) & (...) & 1.9 \\
2012 Jun 09 & 16:45:00 & M1.9 & 200 & 193 & 18.5 & 16:50:17 & 2.5 & 16:53:38 & 1.0 & 0.8 & (...) & (...) & (...) \\
2012 Jun 10 & 06:39:00 & M1.3 & 317 & 107 & 17.5 & 06:43:21 & 6.3 & 06:48:38 & 1.0 & 1.1 & (...) & (...) & 0.7 \\
2012 Jun 30 & 12:48:00 & M1.1 & 130 & 64 & 17.3 & 12:51:33 & 6.3 & 12:53:43 & 1.6 & 1.0 & (...) & (...) & 0.2 \\
2012 Jun 30 & 18:26:00 & M1.6 & 185 & 80 & 17.7 & 18:31:08 & 6.3 & 18:34:13 & 1.4 & 1.2 & (...) & (...) & 0.4 \\
2012 Jul 02 & 00:26:00 & M1.1 & 351 & 174 & 17.2 & 00:33:41 & 2.0 & 00:39:33 & 0.8 & 1.7 & (...) & (...) & 0.7 \\
2012 Jul 02 & 19:59:00 & M3.8 & 694 & 186 & 18.9 & 20:04:29 & 2.0 & 20:16:03 & 0.9 & 2.4 & 1.0 & Al\_thick & 5.3 \\
2012 Jul 04 & 14:35:00 & M1.3 & 138 & 84 & 18.6 & 14:39:24 & 7.9 & 14:41:43 & 1.2 & 1.9 & (...) & (...) & (...) \\
2012 Jul 05 & 01:05:00 & M2.5 & 208 & 206 & 19.2 & 01:09:34 & 7.9 & 01:13:03 & 0.6 & 2.0 & (...) & (...) & (...) \\
2012 Jul 05 & 10:44:00 & M1.8 & 150 & 565 & 19.9 & 10:47:23 & 7.9 & 10:49:53 & 0.1 & 1.5 & (...) & (...) & (...) \\
2012 Jul 05 & 13:05:00 & M1.2 & 1010 & 119 & 14.4 & 13:11:03 & 7.9 & 13:27:53 & 0.6 & 1.0 & (...) & (...) & (...) \\
2012 Jul 06 & 01:37:00 & M3.0 & 150 & 52 & 22.0 & 01:39:23 & 2.0 & 01:41:53 & 4.1 & 0.5 & (...) & (...) & 0.8 \\
2012 Jul 06 & 08:17:00 & M1.6 & 187 & 421 & 17.8 & 08:23:06 & 7.9 & 08:26:14 & 0.3 & 1.8 & (...) & (...) & (...) \\
2012 Jul 06 & 23:01:00 & X1.1 & 178 & 239 & 26.7 & 23:07:05 & 2.5 & 23:10:04 & 1.2 & 2.6 & (...) & (...) & (...) \\
2012 Jul 07 & 03:10:00 & M1.2 & 472 & 77 & 20.0 & 03:13:11 & 7.9 & 03:21:04 & 1.1 & 0.2 & (...) & (...) & (...) \\
\end{tabular}
\end{table}

\clearpage
\end{landscape}

\end{document}